\documentclass[11pt,a4paper,]{article}
\usepackage{jinstpub}

\usepackage{amsmath}
\usepackage{ulem}
\usepackage{natbib}
\usepackage{mathtools}
\usepackage{comment}

\usepackage{gensymb}
\usepackage{graphicx}
\usepackage[colorinlistoftodos]{todonotes}

\usepackage[colorlinks=true]{hyperref}

\usepackage{siunitx}
\usepackage{float}
\usepackage{appendix}
\usepackage{wrapfig}

\usepackage{lineno}

\usepackage{notoccite}

\usepackage{textgreek}
\begin{document}
\title{Fluorescence of pyrene-doped polystyrene films from room temperature down to 4~K for wavelength-shifting applications\\}

\author[a]{H.~Benmansour}
\author[a,d]{E.~Ellingwood}
\author[a]{Q.~Hars}
\author[a,1]{P.C.F.~Di~Stefano\note{Corresponding author.}}
\author[b]{D.~Gallacher}
\author[c]{M.~Ku\'zniak}
\author[a,d]{V.~Pereimak}
\author[b]{J.~Anstey}
\author[b]{M.G.~Boulay}
\author[b]{B.~Cai}
\author[b]{S.~Garg}
\author[a,d]{A.~Kemp}
\author[b]{J.~Mason}
\author[a,d]{P.~Skensved}
\author[b]{V.~Strickland}
\author[a,d]{M.~Stringer}
%\author{\textcolor{red}{Full author list TBD}}

\affiliation[a]{Department of Physics, Engineering Physics \& Astronomy, Queen’s University, Kingston, ON, K7L 3N6, Canada}
\affiliation[b]{Department of Physics, Carleton University, Ottawa, K1S 5B6, ON, Canada}
\affiliation[c]{AstroCeNT, Nicolaus Copernicus Astronomical Center, Polish Academy of Sciences, Rektorska 4, 00-614 Warsaw, Poland}
\affiliation[d]{Arthur B. McDonald Canadian Astroparticle Physics Research Institute, Queen’s University, Kingston ON K7L 3N6, Canada}

\emailAdd{distefan@queensu.ca}

\abstract{In liquid argon-based particle detectors, slow wavelength shifters (WLSs) could be used alongside the common, nanosecond scale, WLS tetraphenyl butadiene (TPB) for background mitigation purposes.
At room temperature, pyrene has a moderate fluorescence light yield (LY) and a time constant of the order of hundreds of nanoseconds. In this work, four pyrene-doped polystyrene films with various purities and concentrations were characterized in terms of LY and decay time constants in a range of temperature between 4~K and 300~K under ultraviolet excitation. These films were found to have a LY between 35 and 50\% of that of evaporated TPB. 
All light yields increase when cooling down, while the decays slow down. At room temperature, we observed that pyrene purity is strongly correlated with emission lifetime: highest obtainable purity samples were dominated by decays with emission time constants of $\sim$~250--280~ns, and lower purity samples were dominated by an $\sim$~80~ns component.
One sample was investigated further to better understand the monomer and excimer emissions of pyrene. The excimer-over-monomer intensity ratio decreases when the temperature goes down, with the monomer emission dominating below $\sim 87$~K.}

\keywords{Wavelength shifter, Pyrene, Fluorescence, Liquid argon, Excimer, Monomer}

\maketitle
\clearpage

\section{Introduction}

Noble liquids are used as particle detection mediums due to their high scintillation light yields, excellent radio purity capabilities and particle identification abilities. Experiments using noble liquid detectors include direct dark matter detection~\cite{DEAP2,DARKSIDE50,XENON1T,LUX,XMASS1,PANDAXX}, neutrinoless double-beta decay~\cite{NEXO}, and neutrino observation~\cite{DUNE}.
In such detectors, a particle interaction in the liquid produces scintillation light roughly proportional to the energy deposited. The vacuum ultra-violet (UV) scintillation emission of liquid argon and xenon ($\sim 128$~nm and $\sim 175$~nm respectively) make light detection difficult with standard detectors such as photomultipliers (PMs)~\cite{instruments5010004}.

One common way to deal with this is to use a wavelength shifter (WLS), a material which absorbs the UV light and re-emits it as easier-to-detect visible light.  A standard WLS is 1,1,4,4-tetraphenyl-1,3-butadiene (TPB), characterized by a short, $\sim$~ns, time constant~\cite{Flournoy}, and high conversion efficiency~\cite{lally_uv_1996}.  In certain cases, for instance to identify different regions of the detector~\cite{gallacher_2021_arxiv}, it may be advantageous for the WLS to have significantly longer time constants than TPB and argon, enabling a form of spatial pulse-shape discrimination. A similar approach can also be used in liquid scintillator detectors for separation of Cherenkov and scintillation components~\cite{BILLER2020164106} or other purposes~\cite{HAMEL2021118021}.

In this work, we study the fluorescence light yield and decay profiles of four pyrene-doped polystyrene (PS) coatings with diverse pyrene purities and concentrations, excited by a 285~nm LED at temperatures between 4~K and 300~K, including the 87~K boiling point of liquid argon. 
Fluorescence yield and decay profiles were both determined from time-resolved measurements.  One sample was investigated further with spectral measurements to characterize the monomer and excimer responses.

\section{\label{sec:pyrene_fluo}Wavelength Shifters, Excimers and Monomers}

In particle detectors using noble liquids such as argon, a WLS coating is generally applied between the liquid and the light detectors.    WLS are  materials that absorb the UV light and re-emit it by fluorescence in the visible spectrum.  As noble elements are  liquids at low temperatures (87~K for argon), it is important to characterize the WLS in such conditions.

Pyrene (C${}_{16}$H${}_{10}$) is a polycyclic aromatic hydrocarbon  consisting of four fused benzene rings. Two adjacent molecules of pyrene (monomers) can bond together forming a dimer with an interplanar distance of 3.5~\AA~\cite{Birks}.   Both monomers and dimers can be excited; the latter are called excimers.  As a pure crystal, pyrene has a high vapor pressure of $6 \times 10^{-6}$~mbar~\cite{sonnefeld_dynamic_1983} making it unsuitable for the vacuum requirements of many wavelength-shifting setups.   It is therefore often used as a dopant in a more stable  matrix of a material like poly(methyl methacrylate) (PMMA, a.k.a. acrylic)~\cite{captalk,itaya_interfacial_1990} or polystyrene (PS)~\cite{itaya_interfacial_1990}.
Room-temperature fluorescence spectra of pyrene in PS  films with different pyrene concentrations exhibit two distinct, but broad, emission features, from the monomer ($\sim$~395~nm), and excimer states ($\sim$~470~nm), of pyrene~\cite{Johnson}.  The  amount of excimer emission increases with pyrene concentration~\cite{Johnson}.
At room temperature, fluorescence decay of diluted monomer happens on a scale of $\sim$~200~ns~\cite{Johnson} --- of interest here, since it is significantly slower than TPB and the singlet lifetime of liquid argon.

\begin{figure}[h]
\centering
\includegraphics[width=0.8\textwidth,trim=0pt 190pt 0 190pt]{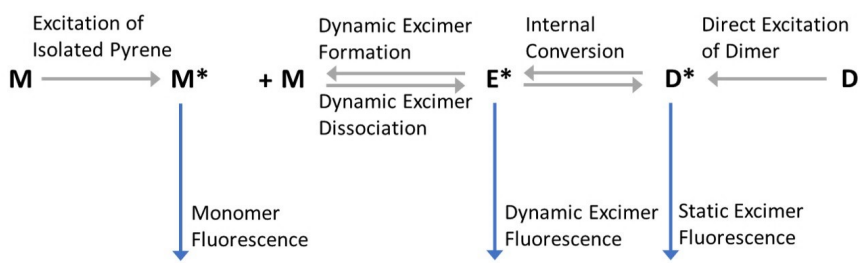}
\caption{\label{fig:levels}Pyrene fluorescence diagram. M is a ground state pyrene, M* is a pyrene monomer in the first excited state. E* is the dynamic excimer formed by (M*+M), D is the associative ground state pyrene dimer and D* is the dimer in the first excited singlet state. Grey arrows relate to the pyrene molecule processes and blue relate to the fluorescence. Adapted from~\cite{Winnik}.}
\end{figure}

Fluorescence of pyrene in solid matrices involves three mechanisms~\cite{Winnik}.  Monomer emission occurs when an isolated pyrene molecule absorbs light, then de-excites radiatively to the ground state.  
Dynamic excimer emission takes place when an exciton migrates from an excited monomer to a preassociated dimer~\cite{Rabek}, creating an excimer which de-excites radiatively.
Formation of  dynamic excimer typically takes place on timescales of nanoseconds or more.
Static excimer emission happens when a preassociated dimer is excited, then de-excites radiatively. Due to their proximity, the molecules in the dimer have perturbed absorption and excitation spectra.  Static formation takes place on  picosecond timescales.  Conversion can occur between the  processes described above.
Mechanisms of monomer and excimer fluorescence are summarized in Fig.~\ref{fig:levels}.
While the probabilities of radiative decay do not depend on temperature, temperature dependent non-radiative decays can also take place.  The probability of these tends to decrease as temperature falls, contributing to slower decays and generally higher light yields~\cite{TAKAHASHI_pyrene}.

\section{\label{sec:setup}Equipment and Setup Details}

The pyrene fluorescence studies involve data  from two different setups. Both  feature  acrylic substrates coated with pyrene-polystyrene films installed in an optical cryostat~\cite{cryostat} to make measurements at  temperatures from 300~K down to 4~K under UV excitation.  Measurements were taken sequentially while cooling to avoid thermoluminescence, with cross-check measurements performed while warming.  Cryostat pressure was $\lesssim 10^{-6}$~mbar.

The time-resolved measurements of the pyrene sample are used to study the fluorescence light yield (number of photons emitted by fluorescence for a given excitation) and decay time of the pyrene coating. This setup features a fast pulsing LED exciting the sample and a PMT to collect the pyrene fluorescence light. The spectral measurement uses the same LED, but with a constant brightness, to excite the sample and a spectrometer to analyze the pyrene fluorescence spectrum.

\begin{figure}[h]
\centering
\includegraphics[width=0.55\textwidth,trim=30pt 170pt 30pt 170pt]{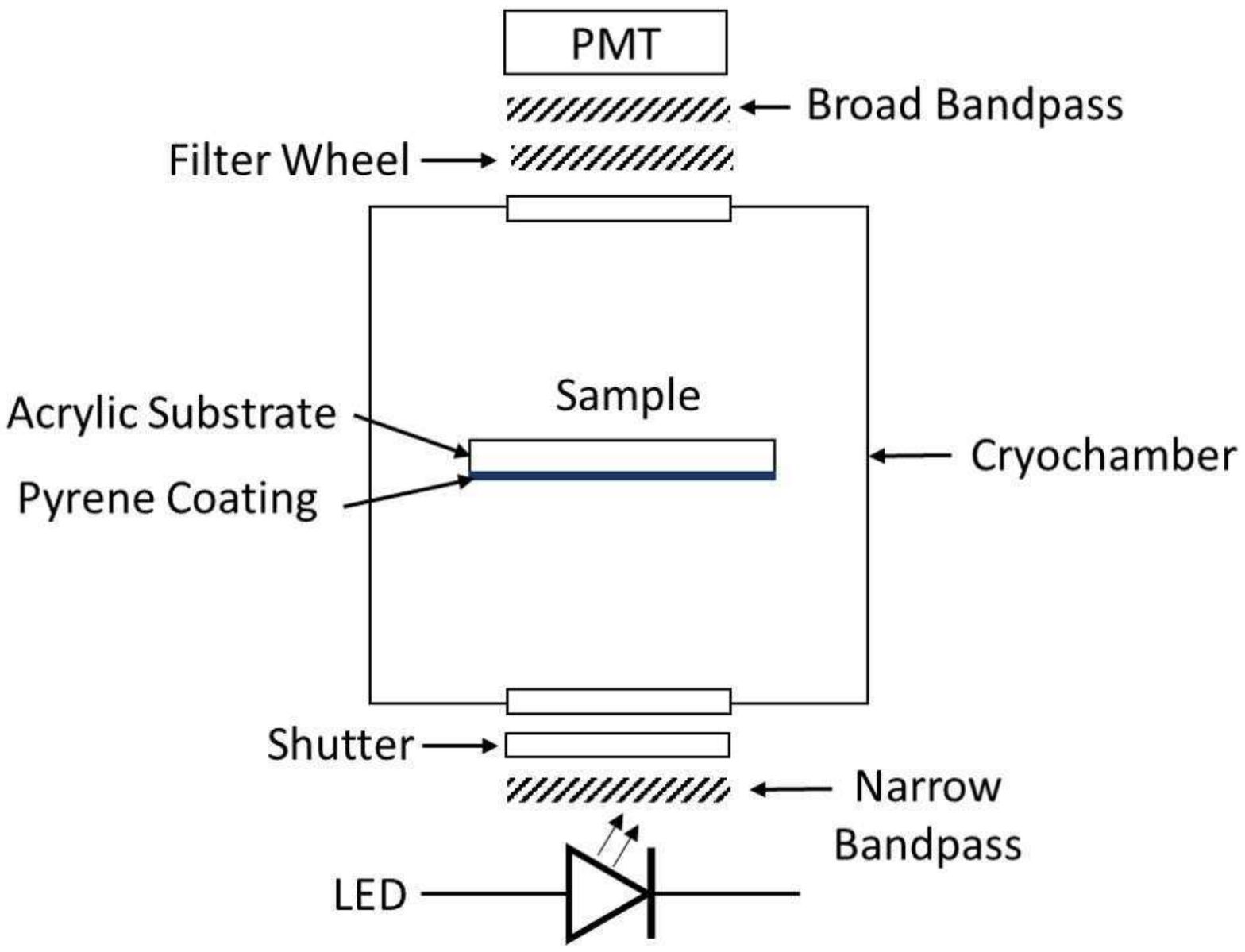}
\hfill
\includegraphics[width=0.30\textwidth]{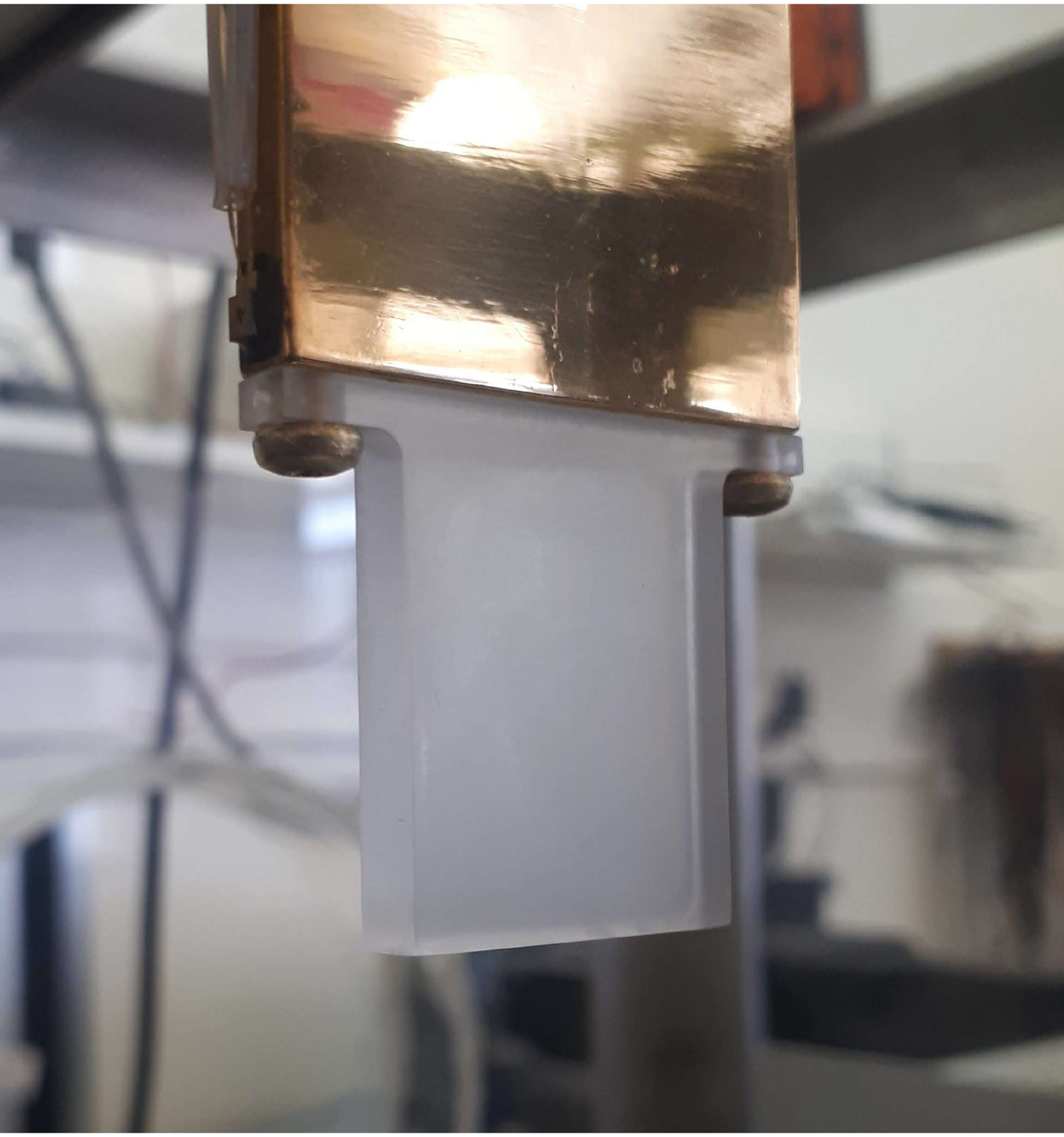}
\caption{\label{fig:setup}Left: Setup for time-resolved measurements. 285~nm LED light interacts with the pyrene and the acrylic from the sample. The resulting fluorescent light from the sample  passes through a filters to select all or parts of the spectrum before being detected on the PMT. 
Adapted from~\cite{jasmine_lidine}. 
Right: Image of an acrylic sample attached to the cold finger of the optical cryostat. The cryostat thermometer is visible on the left side of the cold finger.}
\end{figure}

\subsection{Samples}

The samples used in this study were made at Carleton University and consist of a pyrene polystyrene mix applied as a film on Reynolds Polymer Technology (RPT) acrylic substrates that were machined to fit in the optical cryostat as shown in Fig.~\ref{fig:setup}. Details of pyrene coating preparation can be found elsewhere~\cite{gallacher_2021_arxiv}.  Information on the RPT acrylic can be found in~\cite{jasmine_lidine}.  

A total of four different pyrene-polystyrene mixes are studied in this paper. There are three samples with a 15\% pyrene concentration by weight (equivalent to $7.78\times10^{-4}$ mol/$\mbox{cm}^3$) but three different fluorescence grades (98\%, 99\% and 99.9\% purity), called P1598, P1599 and P15 respectively. There is one sample with a 12\% pyrene concentration in weight (equivalent to $6.22\times10^{-4}$ mol/$\mbox{cm}^3$) and a 99.9\% fluorescence grade called P12. The pyrene-polystyrene coating was applied on the acrylic substrate with a syringe for P12 and P15, and with a nylon bristled brush for P1598 and P1599.

This report will compare the results from these four samples. An identical substrate coated with evaporated TPB provides a fifth sample for reference. The full analysis of TPB will be presented elsewhere~\cite{qTPB}.

\subsection{\label{sec:setup_time}Time-Resolved Mode}

The setup used for the time-resolved measurements in presented in Fig.~\ref{fig:setup}.
The excitation is provided by a 285~nm LED. A function generator produces a 2~V amplitude 50~Hz square wave. This signal is fed into the pulser system which consists of a  Kapunstinsky fast-timing pulser circuit~\cite{Kapustinsky} that powers the UV LED. This system produces a 285~nm pulse which has a FWHM of $\sim 6$~ns. Between the LED and the entry window of the cryostat there is a TECHSPEC Hard Coated OD~4.0 10~nm bandpass filter with a central wavelength at 280~nm and a 10~nm FWHM. This filter ensures that the sample is only excited by the wavelength of interest from the LED. 

The luminescence of the pulser system can be dependent on the ambient temperature of the room. A  W1209 temperature controller was used to stabilize the temperature of this system by heating it with a resistor  attached to the outside of the metal box housing the LED and the Kapustinsky circuit. This kept the circuit and LED at (29.0$\pm$0.1)$^{\circ}$C.

The 285~nm LED pulses interact with the pyrene coating producing a distinct fluorescence spectrum with a light yield and pulse decay timing that should depend on the temperature of the sample inside the cryostat.  The light then goes through the acrylic substrate, which cuts off any remaining UV component.  The optical ordering of UV light, pyrene-doped coating, acrylic, light detection, is similar to that of an argon particle detector like DEAP, made of acrylic, with the inner surface coated with pyrene-doped polystyrene and the light detectors on the other side~\cite{gallacher_2021_arxiv}.

\begin{figure}[h]
\centering
\includegraphics[width=0.999\textwidth]{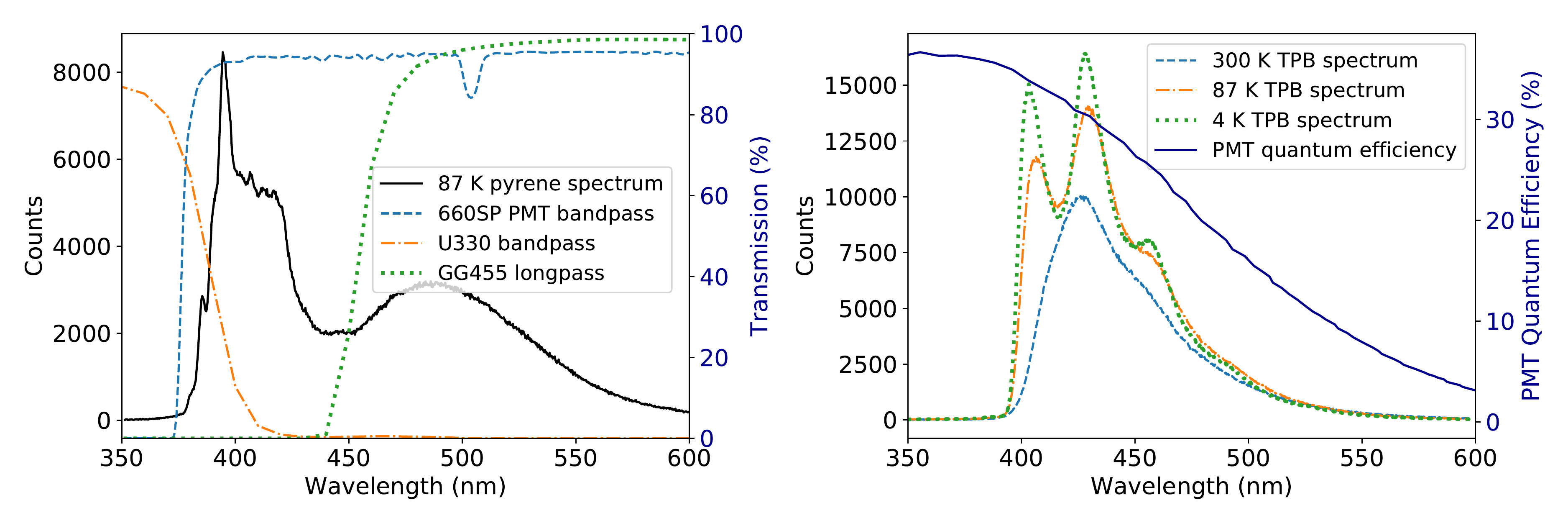}
\caption{Left: Filters used in time resolved measurements. 660SP Rapid Edge bandpass filter between the cryostat exit window and the PMT, U330 bandpass filter for monomer measurements, GG455 filter for excimer. The P15 spectrum at 87 K with only the PMT bandpass applied is also shown to demonstrate where these filters cut on the spectrum. Right: TPB spectra at multiple temperatures~\cite{jasmine_lidine}, compared to the PMT quantum efficiency, as a function of wavelength~\cite{hamamatsu_PMT_QE}.}
\label{fig:all_filters}
\end{figure}

The light emitted by the sample passes through the exit window of the cryochamber and then goes through a 660SP Rapid Edge bandpass filter from Omega Optical which accepts light between 375~nm and 660~nm in order to reject any remaining 285~nm LED light. 
A motorized filter wheel was installed between the UV rejection filter and the PMT to switch between filters that could separate the monomer and excimer components of the pyrene fluorescence emission.
The excimer is studied using a Schott GG455 longpass filter with a cut-on wavelength of 455~nm. The monomer part of the spectrum is studied using a Hoya U330 bandpass filter with a 330~nm central wavelength and a 140~nm FWHM. The transmission spectrum for these two filters and the broad bandpass spectrum are shown in Fig.~\ref{fig:all_filters}.

A Hamamatsu R6095-100 photomultiplier tube (PMT) with a super-bialkali photocathode and borosilicate glass is used to collect the fluorescence light from the pyrene which is powered by a CAEN high voltage power supply and recorded using a digitizer. This study uses a National Instruments PXIe-5160 digitizer which has a 10-bit analog input resolution with a 2.5~GS/s sampling rate.
The 50~Hz function generator pulse that is fed into the Kapustinsky circuit is also used as the trigger for the digitizer while the PMT output is fed into one of the digitizer channels.  The PMT pulse is recorded for 12~$\mu$s sampled every 0.4~ns.  The first 1.2~$\mu$s of the record are pretrigger.  A run at a specified temperature consists of 45~000 of these 12~$\mu$s events.

 \subsection{\label{sec:spectra_setup}Spectral Mode}

The basic setup of this experiment is similar to the time-resolved measurements in its use of the 285~nm LED and the cryostat. The spectra were taken at seven temperatures: 300~K, 250~K, 163~K, 87~K, 77~K, 27~K and 4~K.
The two main differences between the time-resolved and spectrometer setup were the LED mode and that the PMT was replaced with a spectrometer.
The 285~nm LED was run in DC mode to provide a continuous light source as opposed to a pulsed source. To do this the Kapustinsky circuit was bypassed and the LED was provided with a continuous power input.

This experiment uses a Horiba microHR spectrometer with a Horiba Symphony II liquid-nitrogen cooled open electrode 2D CCD array. The spectrometer and the cryostat are connected via an optical fiber at the same position on the cryostat as the PMT was for the time-resolved measurements. The spectrometer settings, including the filter, wavelength range and exposure, and data acquisition are controlled by the SynerJY program on the DAQ computer. The  spectrum window was from 350--650~nm and the exposure was set to 10~s. For each spectrum, an additional exposure with the system shielded from light was subtracted to remove dark counts.

\section{Analysis Techniques}

\subsection{Data Reduction}

The digitizer is triggered and synchronized with the LED pulser so there are no spurious pulses in the data set and no cuts have been applied to the data. The data are reduced by two different methods:
\begin{itemize}
    \item An average pulse is calculated for all the events in a run; the baseline is determined from the pretrigger and subtracted from the average.
    \item Reduced quantities are calculated for individual pulses; they include the baseline and its sample-by-sample standard deviation, and pulse integral between user-specified bounds.
\end{itemize}
Data have been analyzed and plotted using the NumPy~\cite{harris2020array} and Matplotlib~\cite{Hunter:2007} python libraries.

\subsection{\label{sec:noise}LED-Induced Noise}
In all time-resolved measurements, an electronic noise oscillation around the baseline was observed which is due to the pulser circuit.  To better understand its characteristics and mitigate its effects, there were dedicated noise runs at at least one temperature, usually 300~K, for each study of a sample. The noise run was taken with an identical procedure to the other time-resolved measurements except that the shutter between the LED and cryochamber was closed. It was then possible to observe noise-only events without the light emitted by the sample.

\begin{figure}[h]
\centering
\includegraphics[width=0.8\textwidth]{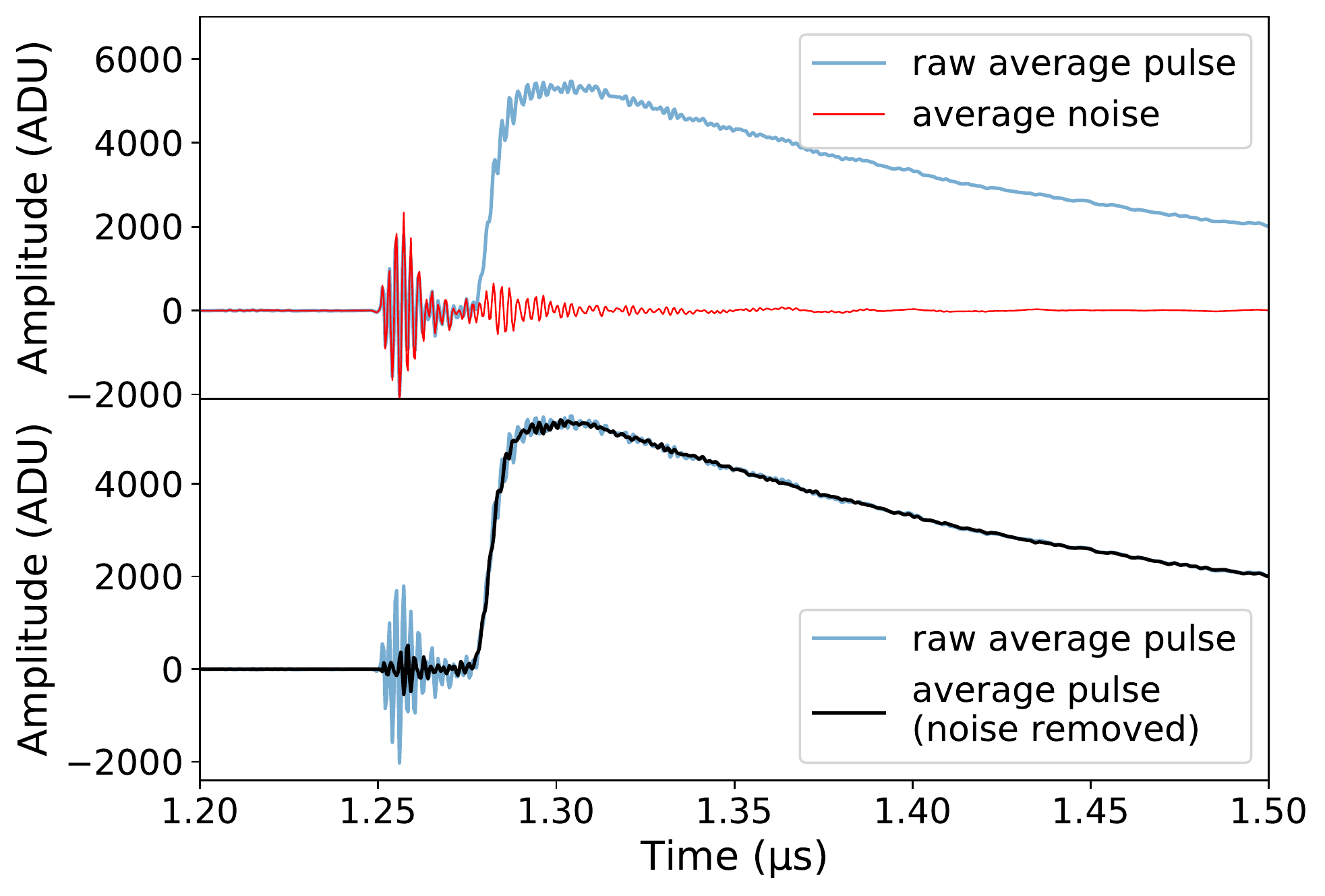}
\caption{For the analysis, the average noise (red) is removed from the raw average pulse (blue). The average pulse (black) obtained is used to do the fits.}
\label{fig:noise}
\end{figure}

The red curve in Fig.~\ref{fig:noise} (top) is an average of the 45~000 noise events. We notice that there is a consistent pattern of fast oscillations with a frequency of several hundreds MHz. This consistency, both in shape and timing, allows us to mitigate the noise by simply subtracting the average pulse shape of a noise-only run from the average shape of any run. The noise and its fluctuations are accounted for in the $\chi^2$ calculation used for pulse fitting as explained in Sec.~\ref{sec:adhoc_fits}.

\subsection{SPE Analysis} 

\begin{figure}[h]
\includegraphics[width=0.99\textwidth]{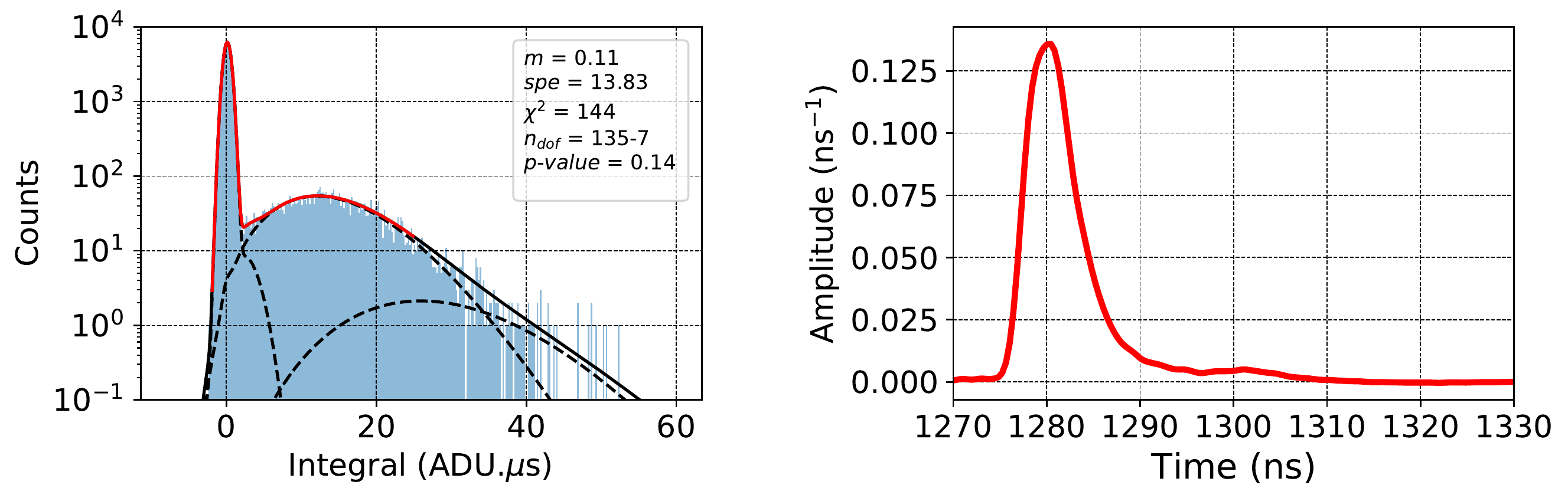}
\caption{\label{fig:SPE_shape}Data obtained with no sample in an empty cryostat.  Left: Fitted single photoelectron charge distribution. 
Average number of photoelectrons per LED pulse is $m=0.11$; average SPE integral is $13.83 \pm 0.04 \mbox{ ADU}\cdot\mu\mbox{s}$.
Right: Normalized shape of LED pulse and PM response.}
\end{figure}

This analysis determines  the number of detected photoelectrons as a proxy for light yield.
For this,  the average single photoelectron (SPE) integral has to be estimated. 
SPE integral distributions are obtained by reducing the LED voltage to a level where the PMT detects one photoelectron or less for each event. The integration time for each pulse is set to 60~ns. The distribution of these integrals was then fitted with a model describing the photomultiplier response~\cite{CHIRIKOVZORIN2001310}, as shown in Fig.~\ref{fig:SPE_shape}. This SPE integral distribution corresponds to an empty cryostat configuration, i.e. to the LED barely illuminating the PMT with no sample in between. %
The reference SPE integral value corresponds to the average integral of a single photoelectron. This property is characteristic of the PMT and high voltage used, and also depends on the vertical range used on the DAQ window.

For this PMT, several SPE runs were carried out with pyrene samples at different temperatures in addition to the empty-cryostat run in Fig.~\ref{fig:SPE_shape}. The average SPE integral value was found to be equal to $(13.80 \pm 0.04)~\mbox{ADU} \cdot\mu \mbox{s}$, using a vertical range $\pm~0.25~\mbox{V}$ for the DAQ. The integral of a given pulse can then be converted into a number of photoelectrons dividing by this SPE integral value.

Lastly, Fig.~\ref{fig:SPE_shape} illustrates the average shape of the LED pulse and PMT response, obtained in  the empty cryostat,  but with a higher LED intensity to minimize the electronic noise.  After normalization, this is used as the instrumental component during pulse deconvolution (Sec.~\ref{sec:TC_analysis}).
Taking this pulse as a proxy for the noisier actual SPE pulses, the shape justifies the 60~ns integration window.

\subsection{\label{sec:LY_fit}Fluorescence Yield}

\begin{figure}[h]
\centering
\includegraphics[width=0.99\textwidth]{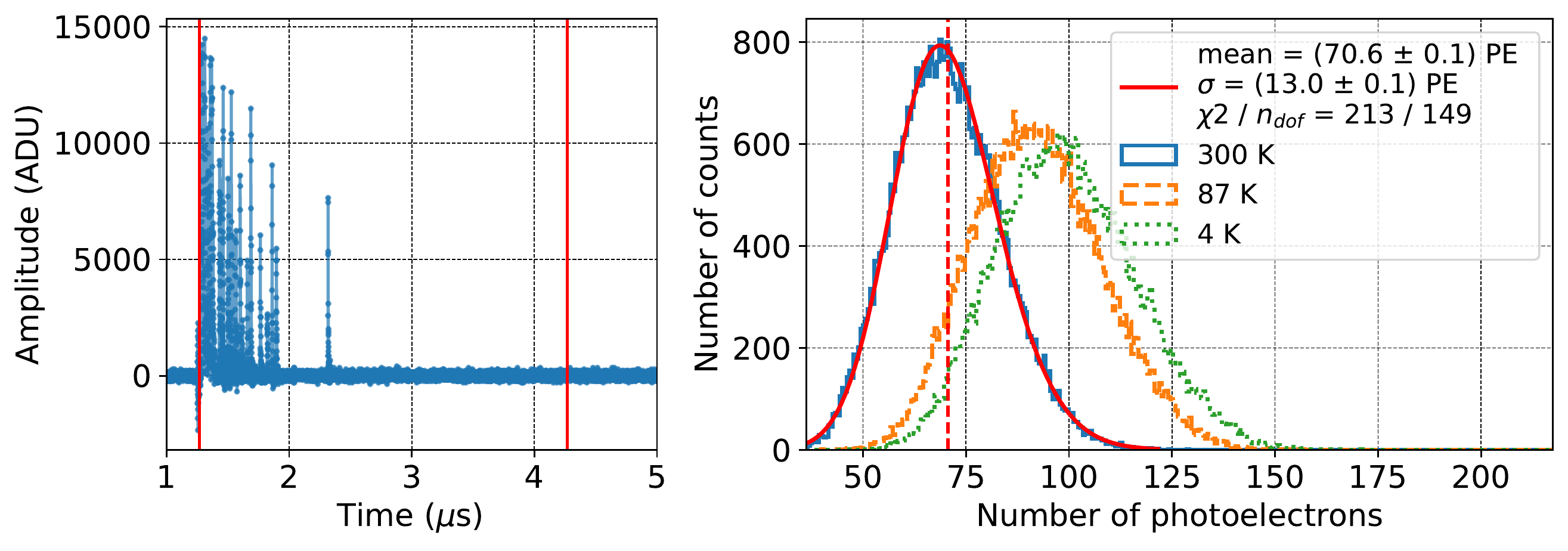}
\caption{\label{fig:integral_window}Left: an event from a pyrene run with two red vertical lines (at 1270~ns and 4270~ns) representing the lower and upper limits of the integration window. Individual photoelectrons are clearly visible.  Right: event integral distribution for three runs at different temperatures. The blue distribution  (300~K) is fitted with a skew normal distribution as an example. The mean of this distribution provided light yield for the sample at that temperature.}
\end{figure}

To estimate the light yield in each set of conditions, we calculate the integral of each event trace in a run by summing the amplitudes in a given window and weighting by the sampling time.  The result is expressed as a number of detected photoelectrons, a proxy for the light yield since the excitation is the same in all runs.
A 3~$\mu \mbox{s}$ window is chosen containing more than 99.9\% of the overall amount of light, as shown in Fig.~\ref{fig:integral_window}.
This figure also shows the distribution of integrals for all 45~000 events in a run for a pyrene sample at different temperatures.

The average detected light yield is then obtained by fitting these distributions with a skew normal function~\cite{skew_normal}. A fit example is shown in Fig.~\ref{fig:integral_window}. It is carried out by minimizing a Baker-Cousins likelihood~\cite{BC}. The mean is represented by a vertical dashed red line. The fit of each distribution allows us to calculate the average detected fluorescence yield of the pyrene sample at each temperature. Comparison between temperatures tells us how the absolute light yield changes with temperature.  Using the same method for the TPB sample, we can access the relative light yield of the sample.

\subsection{\label{sec:TC_analysis}Fluorescence Profile Fitting}

One objective of this report is to explore the fluorescence decay times of the four pyrene samples at different temperatures. This requires fitting the average pulse shape with the right physically motivated mathematical model. Fits were carried out in a $[1285~\mbox{ns} - 5000~\mbox{ns}]$ window containing at least 99.9\% of the light detected in the event. 
Fits of the data ($d_i$ photoelectrons at sample $i$) were carried with a model consisting of the electronic noise ($z_i$ photoelectrons at sample $i$, determined with the shutter closed as in Sec.~\ref{sec:noise}) and the convolved ideal pulse shape ($\nu_i$ photoelectrons at sample $i$) as described in the next paragraph.
Statistical fluctuations of the number of photoelectrons (standard deviation $\sigma_i$) dominate the electronic fluctuations ($\sigma'_i$) as seen in Fig.~\ref{fig:integral_window}.
Fits are carried out by minimizing $\chi^2 = \Sigma_i   \frac{(d_i-z_i-\nu_i)^2}{\sigma_i^2 + \sigma^{'2}_i}$.

The fluorescence profiles $I(t)$ (in photoelectrons/second) can be modelled by the convolution of the excitation pulse $E(t)$ (in $s^{-1}$) and the true decay model $i(t)$ (in photoelectrons/second): $I(t)=E(t) \ast i(t)$. Before fitting, the noise is subtracted from the pyrene average trace in accordance with Sec.~\ref{sec:noise}. The excitation pulse $E(t)$ is obtained by having the LED shining on the PMT directly. The pulse obtained then undergoes noise subtraction and is normalized to an integral of 1, see Fig.~\ref{fig:SPE_shape}.

\subsubsection{\label{sec:adhoc_fits}Fits to Full Pulseshape}

\begin{figure}[h]
  \begin{center}
    \includegraphics[width=\textwidth]{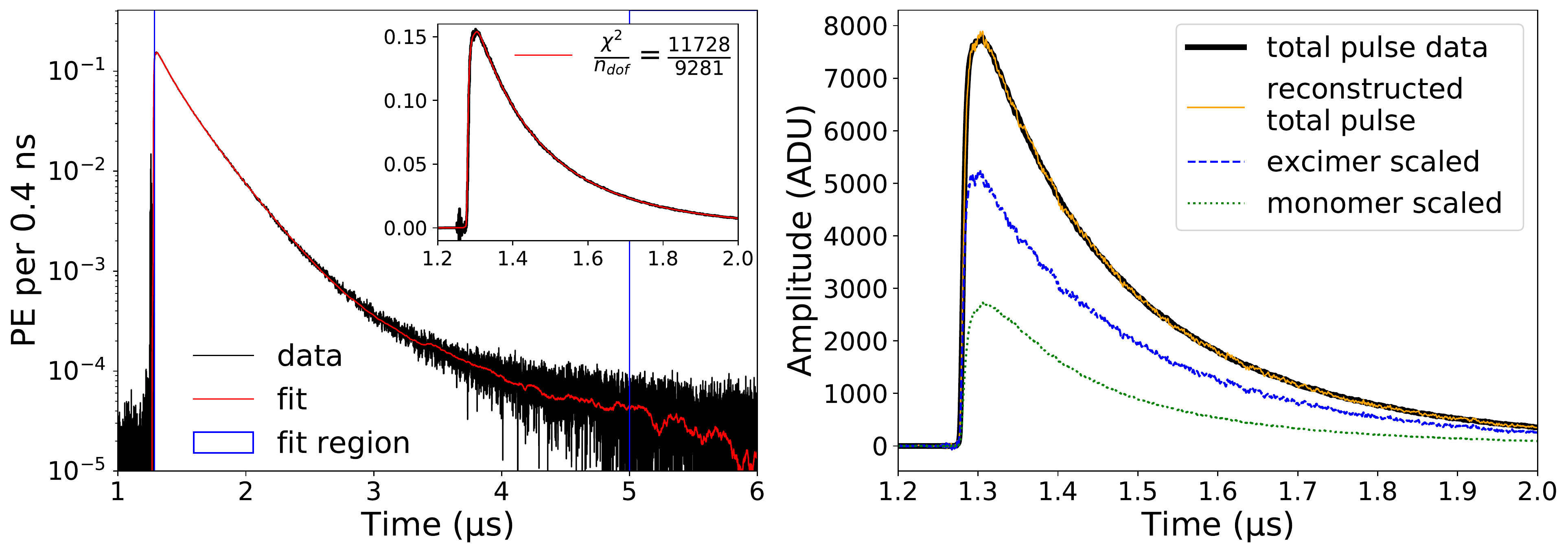}
  \end{center}
  \caption{Left: Fit of P15 fluorescence profile at 87~K using Eq.~\ref{eq:adhoc}. (zoom on  main feature of  pulse and full  window in log scale).  Right: Pulse reconstruction from monomer and excimer profiles at 87~K. The monomer (green) and excimer (blue) contributions are scaled so that the sum (yellow) is the closest fit to the total pulse data (black).}
  \label{fig:adhoc_ex}
\end{figure}

The whole pyrene fluorescence profiles for all four samples were fitted using a sum of three exponentials: two  decays and one  rise. This ad-hoc model allows us to compare the time constants involved in the various pyrene fluorescence profiles of each sample. The model is presented in Eq.~\ref{eq:adhoc} with $N_k\geq0$ and $\tau_k\geq0$:
\begin{equation}\label{eq:adhoc}
i(t)=- \frac{N_{rise}}{\tau_{rise}} e^{-\tfrac{t}{\tau_{rise}}} + \frac{N_1}{\tau_1} e^{-\tfrac{t}{\tau_1}} + \frac{N_2}{\tau_2} e^{-\tfrac{t}{\tau_2}}
\end{equation}
A fit using this model on the whole fluorescence profile of P15 at 87~K is shown in Fig.~\ref{fig:adhoc_ex}.
Errors of $\pm 5\%$  were attributed to the values obtained from the fit to account for the systematic and statistical errors from the fit.
The contribution to the fluorescence light from the different components can be calculated by integrating the  model: $N_{tot} =  \int i(t) dt = N_1 + N_2 - N_{rise}$.  The fractional contribution of each of the three components is :
\begin{equation}\label{eq:Fi}
F_{rise}= \frac{N_{rise}}{N_{tot}} ~~~;~~ \\
F_1= \frac{N_1}{N_{tot}}~~~;~~ \\
F_2= \frac{N_2}{N_{tot}}.
\end{equation}

\subsubsection{\label{sec:separation}Monomer and Excimer Separation}

%%%%%%%%%%%%%%

The pyrene fluorescence is made of two distinct emissions: monomer and excimer. We have observed them separately using a 330~nm bandpass filter and a 455~nm longpass filter (more details in Sec.~\ref{sec:setup}). A linear combination of them matches the full pulse shape, as shown in Fig.~\ref{fig:adhoc_ex} for 87~K. This reconstruction allows us to determine the proportion of monomer and excimer in the total light emitted by the sample at each temperature. To find out the fluorescence decay times, we separately fit the monomer-only and the excimer-only fluorescence profiles.

\begin{figure}[h]
    \centering
    \includegraphics[width=\textwidth]{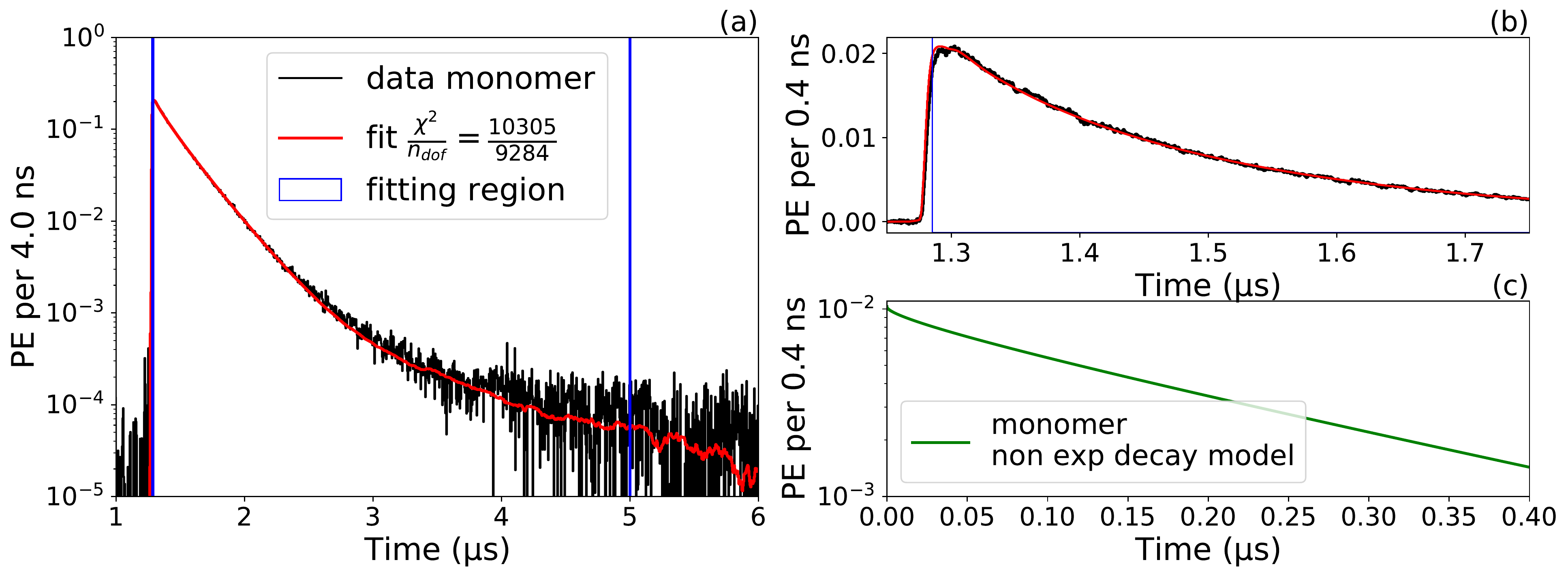}
    \includegraphics[width=\textwidth]{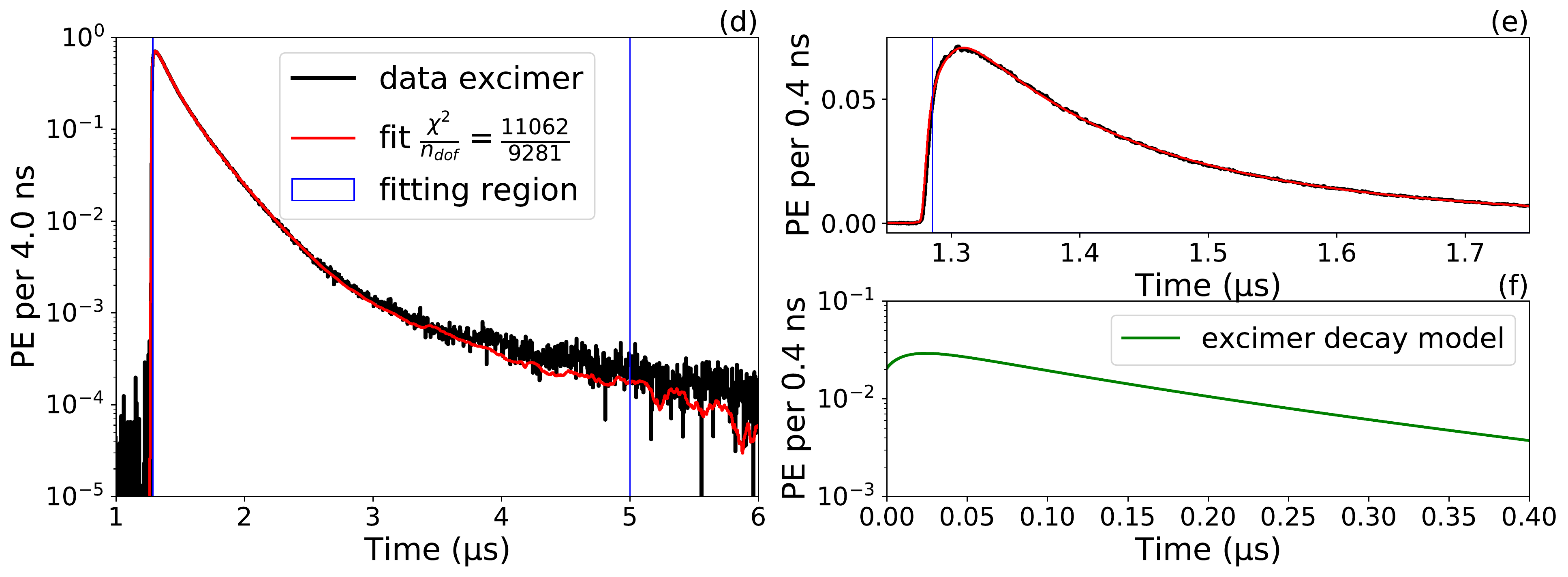}
    \caption{\label{fig:filter_fit} (a): Monomer fit to 87~K data on the full fit window in log scale using the model presented in Eq.~\ref{eq:monomer_fit}. (b): Zoom on fit. (c): and monomer decay model. Bottom three figures (d), (e) and (f) are the same, but for excimer.  Excimer decay model shows a non-zero initial value corresponding to static excimer, and a subsequent rise, corresponding to  dynamic excimer.}
\end{figure}

The unconvoluted monomer profile for pyrene in PS can be described  as an exponential decay accelerated by excimer formation~\cite{Johnson}:
\begin{equation}\label{eq:monomer_fit}
i_{m}(t) = \frac{N_1'}{\tau_1'} e^{-\tfrac{t}{\tau_1'}-2q\sqrt{\tfrac{t}{\tau_1'}}}
\end{equation}
The first parameter is a decay time $\tau_1'$ corresponding to a pure exponential decay in a dilute monomer system with no excimer formation. The second parameter is dimensionless quantity $q$, the ratio of the dimer concentration to a critical dimer concentration, itself defined as the concentration at which the rate transfer from excited monomer to excimer is equal to the total rate of excited monomer decay by all other channels.  This parameter characterizes the non-exponential nature of the decay caused by excimer formation.  Lastly, the total number of photons emitted is $N_1'\left(1- q e^{q^2}\sqrt{\pi} \left[ 1-{\mbox{erf}}(q) \right] \right)$  (Appendix~\ref{app:integral non exp}).  This  model is used to fit the monomer decay in Fig.~\ref{fig:filter_fit}. 
The  decay models for the monomer and excimer are also shown.

For the excimer-only fluorescence, the decay cannot be described by one single exponential; instead we use two decay constants $\tau_2'$ and $\tau_3'$.  In addition, excimer models introduce two populations of excimer:  dynamic  and  static (Sec.~\ref{sec:pyrene_fluo}).  A rise time ($\tau_{rise}'$) is included in the model to account for the formation of the dynamic excimer. 
The model is:
\begin{equation}\label{eq:excimer_fit}
i_e(t) = - \frac{N_{rise}'}{\tau_{rise}'} e^{-\tfrac{t}{\tau_{rise}'}} + \frac{N_2'}{\tau_2'} e^{-\tfrac{t}{\tau_2'}} + \frac{N_3'}{\tau_3'} e^{-\tfrac{t}{\tau_3'}}
\end{equation}
The fit shows that at $t=0$, $i_e\neq0$, supporting the existence of  instantaneous excimer formation, static excimer. The slower rise in the first 20~ns of the pulse indicates the existence of a delayed excimer formation, corresponding to the dynamic excimer.

The contribution to the fluorescence light from the different components can be calculated by integrating the fitting model over the fitting window (Appendix~\ref{app:integral non exp}):
\begin{equation} \label{eq:LY_mono}
 F_{monomer}= \frac{N_m'}{N_{tot}'}=\frac{N_1'}{N_{tot}'} \left(1- q e^{q^2}\sqrt{\pi}\left(1-\mbox{erf}(q)\right)\right)   
\end{equation}

\begin{equation} \label{eq:LY_exc}
 F_{excimer}= \frac{-N_{rise}'+N_2'+N_3'}{N_{tot}'},  
\end{equation}
with $N_{tot}'=N_m'+N_2'+N_3'-N_{rise}'$.

\subsection{Spectra}
For the spectrometer measurements, the main spectrum was recorded between 350~nm and 650~nm, with 0.3~nm bins. The evolution of the monomer and excimer intensities with temperatures are calculated. No correction was applied to the spectral measurements and the analysis obtained from them. The results are shown in Section~\ref{results:spectrum}.

\section{\label{sec:results}Results}

\subsection{\label{results:spectrum}Temperature Dependent Emission Spectra}

We performed detailed spectral measurements of one sample, P15, to study the contributions of monomer and excimer and their temperature-dependent changes. The spectrum also helps to pick suitable filters to study the monomer and excimer components separately for the time-resolved measurements. 

\begin{figure}[h]
  \begin{center}
    \includegraphics[width=0.80\textwidth]{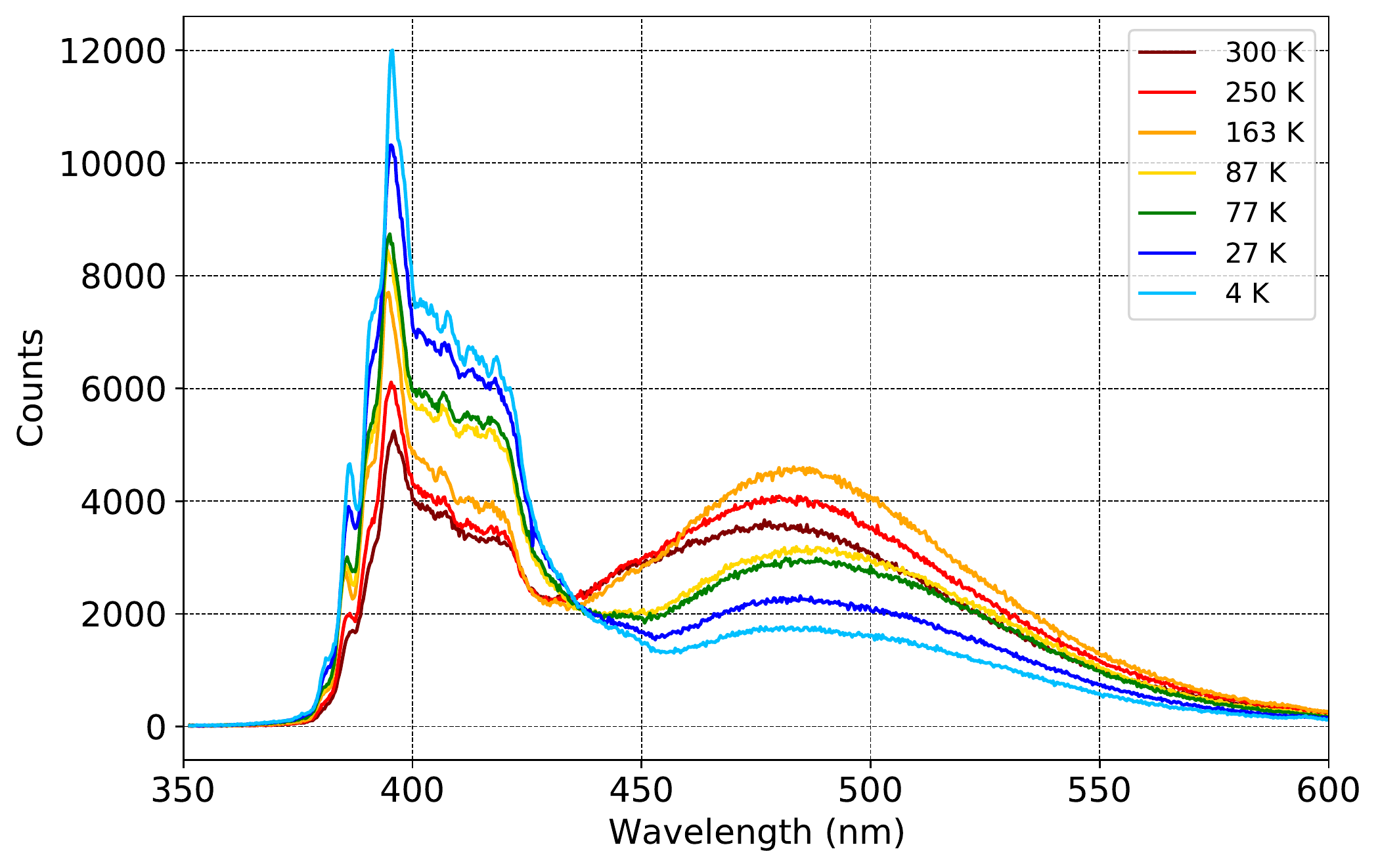}
  \end{center}
  \caption{\label{fig:spectra_T}Spectra of P15 at various temperatures. The main monomer spectrum corresponds to the series of peaks below $\sim 435$~nm  while the excimer corresponds to the broad distribution above about $\sim 435$~nm. The UV-absorbing acrylic substrate explains the sharp drop below 390~nm.}
  
\end{figure}

\begin{figure}[h]
\centering
\includegraphics[width=0.99\textwidth]{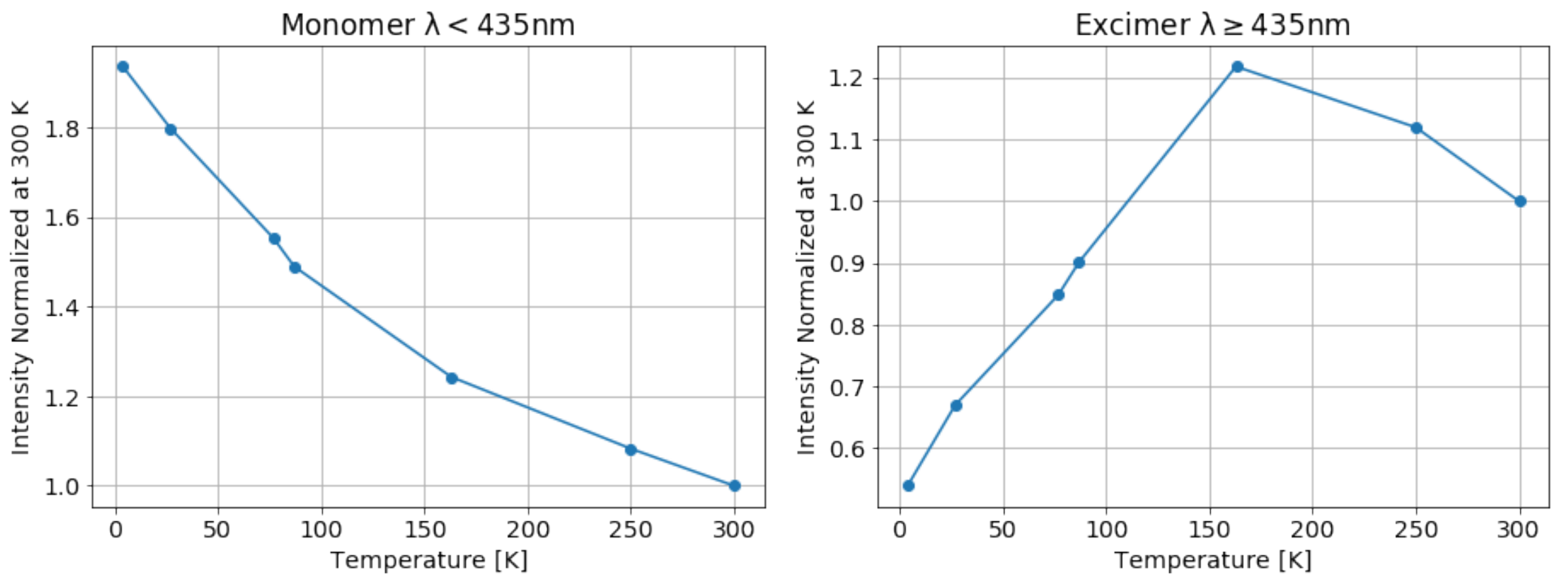}
\caption{\label{fig:monomer_excimer_sum}Monomer (left) and excimer (right) intensity obtained from spectrum integration as a function of P15 temperature. Monomer is defined as the region of the spectrum with wavelengths lower than 435~nm and excimer as the region with higher wavelengths. The intensity of the monomer increases with decreasing temperature, while for the excimer, it increases at first with decreasing temperature and then starts to decrease. }
\end{figure}

Fig.~\ref{fig:spectra_T} shows the raw spectra measured using the cryostat and spectrometer for different sample temperatures. The spectra decrease sharply at wavelengths below 390~nm because of absorption from the samples' acrylic substrate. Based on these spectra, the filters to use in the time resolved measurements were chosen to maximize the light yield from the monomer (330~nm bandpass filter) and excimer components (455~nm longpass filter) in wavelength regions where one of the components dominates. Integrated monomer and excimer contributions are shown in Fig.~\ref{fig:monomer_excimer_sum}.
As the temperature decreases, the monomer contribution increases, while the excimer contribution increases at first and then decreases for temperatures below $\sim$~150~K. 

The filters chosen for the time-resolved measurement monomer and excimer study were based on the measured pyrene fluorescence spectrum. The filter transmission spectra are shown in Fig.~\ref{fig:all_filters}. The 660SP bandpass before the sample is always present during the time-resolved measurements to limit any stray UV LED light from reaching the PMT. To study the whole time-resolved pulse only this filter is present. The monomer is studied using a combination of the U330 bandpass, shown in orange, and the 660SP bandpass. Although the U330 filter cuts out the majority of the peak, the goal was to ensure that the wavelength range limited the broad excimer peak contribution. The excimer is studied using a combination of the GG455 longpass and the 660SP bandpass filters.

\subsection{\label{sec:LY4}Time-Resolved Fluorescence Yield}

\begin{figure}[h]
  \begin{center}
    \includegraphics[width=0.99\textwidth]{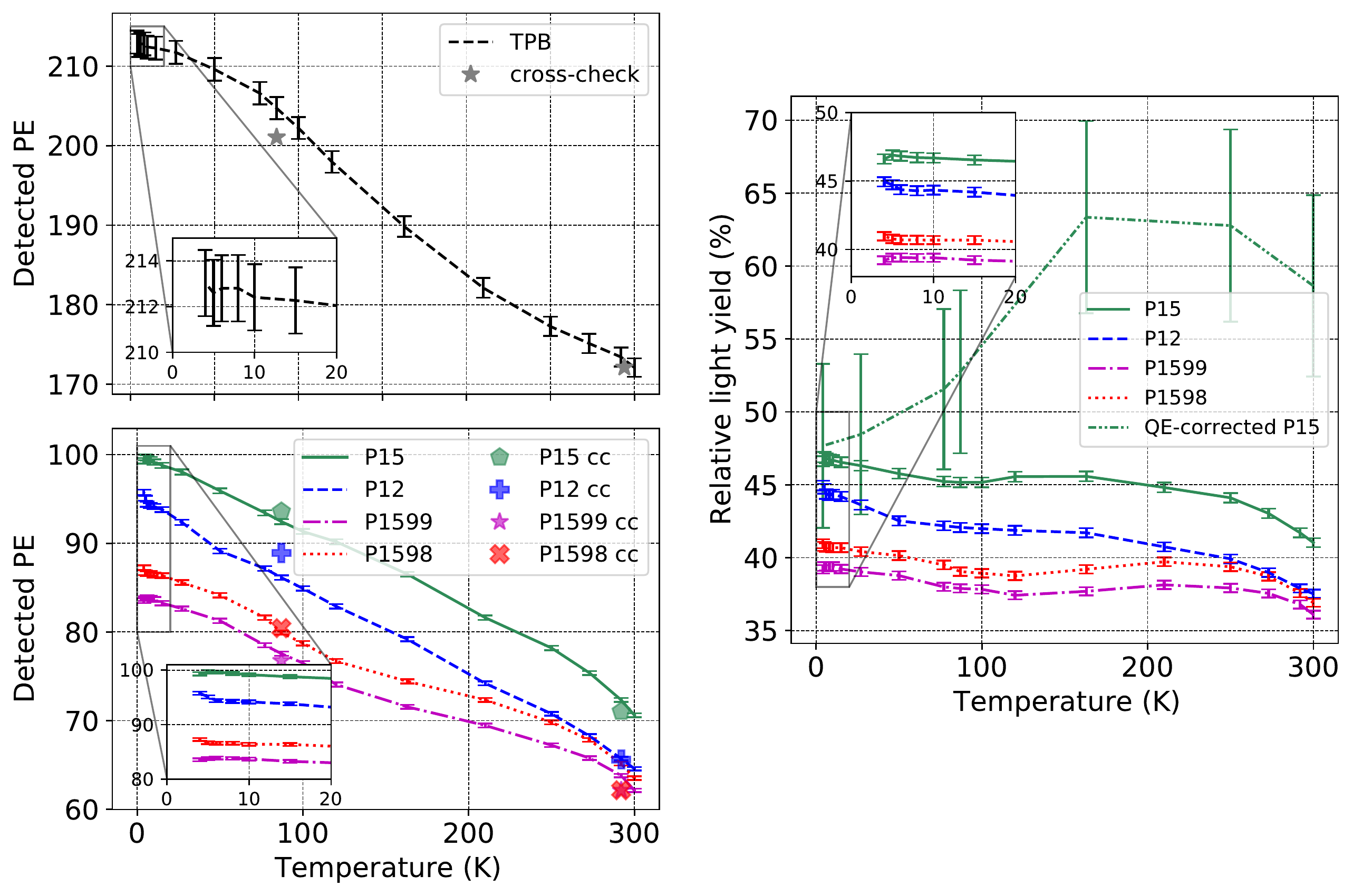}
  \end{center}
  \caption{Left: fluorescence yield of the TPB sample (top) and P15, P12, P1599 and P1598 (bottom) expressed as a number of photoelectrons at different temperatures. Results from cross-check (cc) runs at 87~K and room temperature are also shown. Right: fluorescence yield of the pyrene samples relatively to TPB at different temperatures. Quantum efficiency-corrected P15 relative light yield is also displayed.}
  \label{fig:light_yield_NPE_pyrene}
\end{figure}

The fluorescence yields of four pyrene samples with varying concentrations and purities were studied: P12 (99.9\% purity and 12\% concentration), P15 (99.9\% purity and 15\% concentration), P1599 (99\% purity and 15\% concentration) and P1598 (98\% purity and 15\% concentration). Using the fits of integral distributions presented in Fig.~\ref{fig:integral_window}, the number of photoelectrons detected from each pyrene sample is calculated and compared to the number of photoelectrons detected from TPB~\cite{qTPB} to provide a proxy of the relative light yield. The result is shown in Fig.~\ref{fig:light_yield_NPE_pyrene}.
The amount of light emitted by all pyrene samples compared to TPB is relatively stable and is between 35\% for P1599 at 300~K and 46\% for P15 at 4~K. Going from 300~K to 87~K, the relative light yield for P15 increases by 10\%.

In the case of sample P15, we can use the monomer and excimer data to correct the light yield for PMT quantum efficiency.  The result is overlayed on Fig.~\ref{fig:light_yield_NPE_pyrene}.
This correction increases the relative light yield of P15 with respect to TPB  (eg 60\% vs 41\% at 300~K). The corrected relative light yield falls by 10\% when the temperature decreases from 300~K to 87~K.

\subsection{Fluorescence Decay}

\begin{figure}[h]
  \begin{center}
    \includegraphics[width=0.99\textwidth]{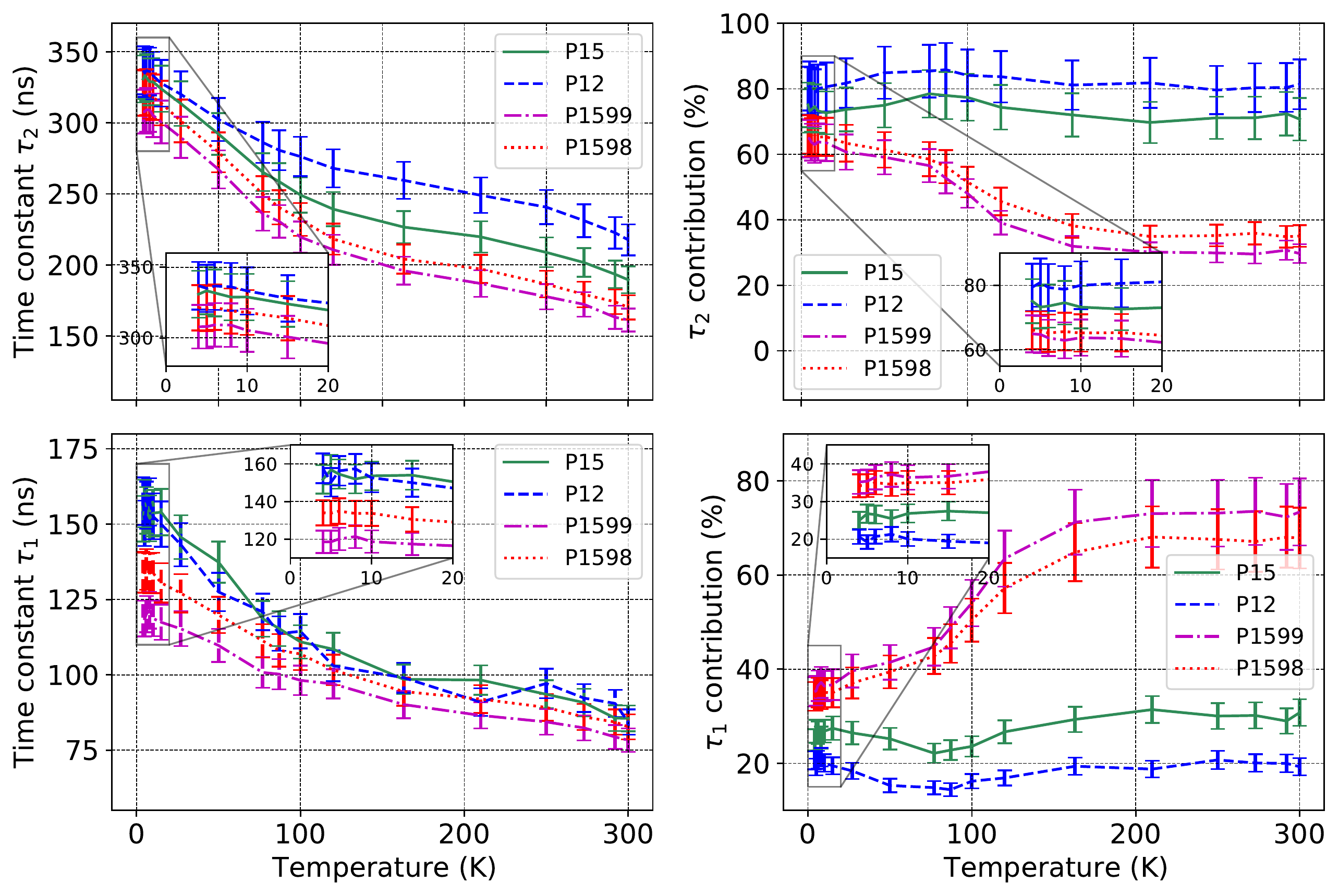}
  \end{center}
  \caption{Left: time constants $\tau_1$ and $\tau_2$ for the four different pyrene samples. Right: contribution of each time constant to the light yield.}
  \label{fig:t1s}
\end{figure}

The fluorescence profiles of the four pyrene samples were fitted using the model from Eq.~\ref{eq:adhoc} in accordance with the method described in Sec.~\ref{sec:adhoc_fits}. Fig.~\ref{fig:t1s} gives the values for the two decay time constants $\tau_1$ and $\tau_2$ for temperatures ranging between 4~K and 300~K for four different pyrene samples: P12 (99.9\% purity and 12\% concentration), P15 (99.9\% purity and 15\% concentration), P1599 (99\% purity and 15\% concentration) and P1598 (98\% purity and 15\% concentration).
The contributions $F_i$ from $\tau_i$ are defined in Eq.~\ref{eq:Fi} (note that $-F_{rise}+F_1+F_2=1$ so $F_1 + F_2 \geq 1$). Because $F_{rise} \leq 4\%$ for all samples at all temperatures, the contribution from the rise component is negligible compared to the other contributions. This contribution is therefore considered as part of the uncertainty on the other contributions.

From these plots it is clear that the concentration and the purity of pyrene seem to have little influence on $\tau_1$ values at higher temperatures. At lower temperatures, the highest purity samples, P12 and P15, show higher values of $\tau_1$. The discrepancy is clearer when looking at the longer time constant, $\tau_2$. Looking at two samples with similar pyrene purities but different pyrene concentrations (P15 and P12), it can be seen that the highest concentration leads to a longer time constant. Looking at three samples with similar pyrene concentrations but different pyrene purities (P15, P1599 and P1598) shows that the highest purity leads to a longer time constant. On the one hand, by comparing P15, P1599 and P1598, it is clear that the lower pyrene purity leads to a much higher contribution from the short time constant $\tau_1$ (70\% vs 30\% at 300~K). On the other hand, a decreased pyrene concentration leads to a higher contribution from the long time constant $\tau_2$ (80\% vs 70\% at 300~K).

\subsection{\label{sec:raw_sep}Fluorescence Decay Times for Excimer and Monomer Emissions of the P15 Sample}

\begin{figure}[h]
  \begin{center}
    \includegraphics[width=0.99\textwidth]{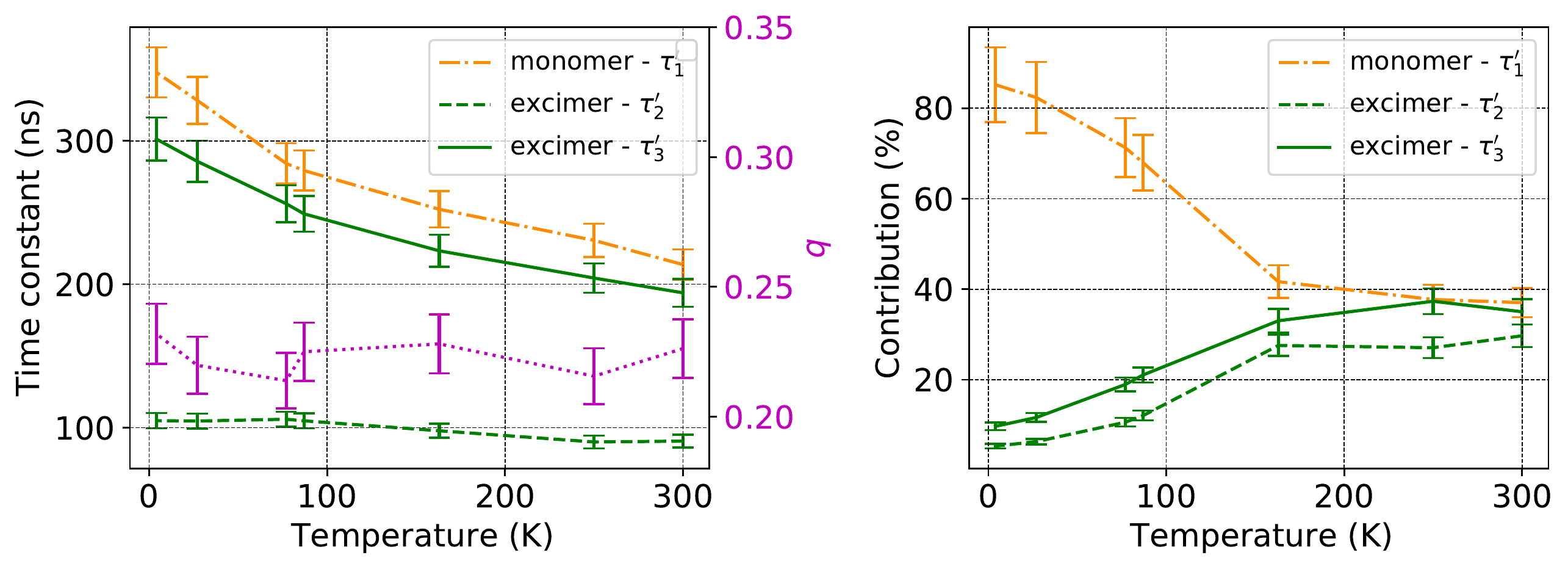}
  \end{center}
  \caption{\label{fig:time_constants}Left: Fluorescence decay times for P15. The monomer profiles are fitted with the model from Eq.~\ref{eq:monomer_fit} and the excimer profiles are fitted with model from Eq.~\ref{eq:excimer_fit}. Parameter $q$ from the model in Eq.~\ref{eq:monomer_fit} is also shown. Right: Contribution to the total light from each component of P15 fluorescence decay. Total excimer represents the sum of the short excimer and the long excimer contribution.}
\end{figure}

The decay profiles are fitted using the models from Eq.~\ref{eq:monomer_fit} and Eq.~\ref{eq:excimer_fit} for the monomer and the excimer profiles respectively. Examples of fit are shown in Fig.~\ref{fig:filter_fit}. Fig.~\ref{fig:time_constants} plots the evolution of the parameter $q$ from the monomer model and of the three fluorescence decay times: one corresponding to the monomer emission, one to the short excimer emission, and one to the long excimer emission. For each component, the contribution to the total light can be determined as a portion of the total light emitted by the sample (Eq.~\ref{eq:Fi}). 
As in the previous section, because $F'_{rise} \leq 4\%$ for all samples at all temperatures, the contribution from the rise component is negligible compared to the other contributions. This contribution is therefore considered as part of the uncertainty on the other contributions. The method is described in Sec.~\ref{sec:separation}.

We notice that $\tau_2'$ is the shortest decay time with values around 100~ns. $\tau_2'$ does not seem to be particularly dependent on the sample temperature. The other decay times all have an increasing trend with  decreasing temperature. The evolution of parameter $q$ from Eq.~\ref{eq:monomer_fit} is also displayed on Fig.~\ref{fig:time_constants}. This parameter from the monomer decay model correlates to monomer conversion into excimers~\cite{Johnson}.

At room temperature, the majority of the light emitted by the sample is due to the excimer emission, but it is equally distributed between the three decay components. When the temperature decreases, the monomer increases its share of the total light while excimer light decreases. At 87~K, the monomer emission represents almost 70\% of the the total light. The decay is therefore dominated by a 280~ns component.

\begin{figure}[h]
  \begin{center}
    \includegraphics[width=0.99\textwidth]{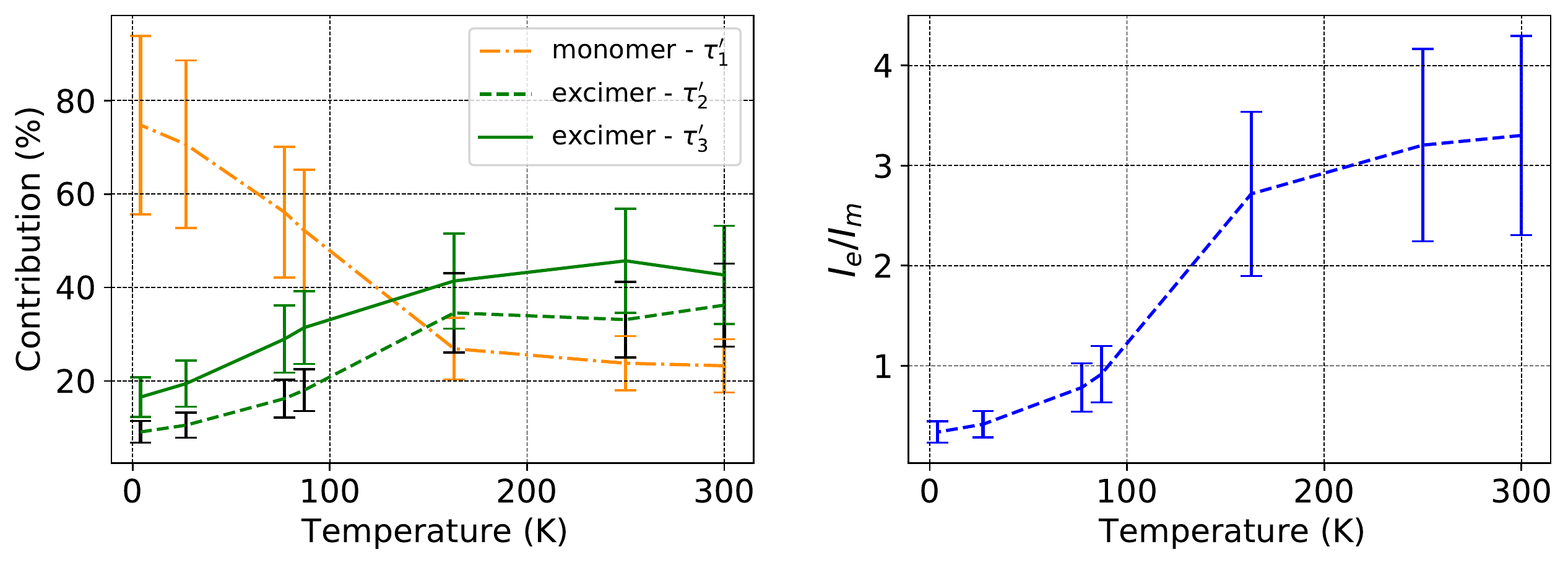}
  \end{center}
  \caption{\label{fig:contributions_corrected}QE-corrected contribution to the total light from each component of P15 fluorescence decay. Total excimer represents the sum of the short excimer and the long excimer contribution. Right: Ratio of the total excimer intensity to the total monomer intensity after PMT QE correction.}
\end{figure}

As discussed in Sec.~\ref{sec:LY4}, the PMT QE differs between the monomer and the excimer emission bands. The contribution to the light presented in Fig.~\ref{fig:time_constants} can therefore be corrected. The QE-corrected contributions are shown in Fig.~\ref{fig:contributions_corrected}.
As expected, the correction gives more importance to the excimer. The monomer is still the dominant component at 87~K after correction but now contributes 50\% of the light at this temperature.

\subsection{Discussion}

Some general trends can be observed in all samples.  The light yields of our pyrene samples all increase when temperature goes down (Fig.~\ref{fig:light_yield_NPE_pyrene}).  This has been observed for pyrene in PMMA as well~\cite{captalk}.  Overall, we attribute this to the shutting down of non-radiative decay channels.  The light yields are between 35\% and 50\% of TPB at all temperatures (Fig.~\ref{fig:light_yield_NPE_pyrene})~\cite{qTPB}.  Regarding the  time constants, all increase during cooling (Fig.~\ref{fig:t1s}), consistent with other measurements of pure pyrene crystals~\cite{Birks} and of pyrene in PMMA~\cite{captalk}, and also consistent with pyrene in solutions at higher temperatures~\cite{HAMEL2021118021}. We attribute this as well to the shutting down of non-radiative decay channels during cooling. Fluorescence profiles were fitted with two decay time constants, a short one between 120~ns and 160~ns at 4~K and a long one between 300~ns and 350~ns at the same temperature (Fig.~\ref{fig:t1s}). This is generally slower than pyrene in PMMA~\cite{captalk}, and significantly slower than the nanosecond timescale of TPB~\cite{qTPB,Flournoy}.
    
As for the effect of concentration and purity, at fixed purity, higher concentration leads to  higher light yield (Fig.~\ref{fig:light_yield_NPE_pyrene}). At a fixed concentration of 15\%, the highest purity sample (P15) has the  highest light yield (Fig.~\ref{fig:light_yield_NPE_pyrene}). This would be consistent with the main obstacle to fluorescence being quenching impurities. However, the trend is reversed between the P1598 and P1599 samples, possibly because the nominal purities are only lower limits on the purity.  The purer pyrenes (P12, P15) tend to have a larger fraction of long decays (Fig.~\ref{fig:t1s}). This is consistent with a smaller number of quenching sites from impurities. However, it is important to note that the two least pure samples coatings were applied using a nylon bristled brush while the two other samples films were applied with a syringe. This might  contribute to the observed behaviours.

Sample P15 was studied in more detail. As the temperature decreases, the monomer contribution to the light yield increases, whereas that of the excimer increases then decreases (Fig.~\ref{fig:monomer_excimer_sum}). Previous work on excimers in pyrene crystals down to 100~K~\cite{Birks} shows the same trend. This behaviour is consistent with the first part of cooling being dominated by the reduction in non-radiative decays, and the second part being dominated by a drop in exciton migration, making it harder for dynamic excimer to form. 
Our excimer-over-monomer intensity ratio decreases when cooling down, as is broadly observed for pyrene in PMMA~\cite{captalk}. Below $\sim 100$~K, the monomer dominates the emission. 

Like the overall time constants, those of the monomer and excimer increase when temperature decreases (Fig.~\ref{fig:time_constants}). Again, it is consistent with non-radiative decay channels shutting down during cooling. We observe no trend in the behavior of $q$ as a function of temperature (Fig.~\ref{fig:time_constants}). This is somewhat surprising, as by definition (Sec.~\ref{sec:separation}), $q$ should depend on non-radiative decay rates.
Alternatively, the long, $3.7 \ \mu$s, window  used in our analysis drowns out the information on $q$. Lastly, our pulse shapes evidence (Fig.~\ref{fig:filter_fit}) the presence of both static and dynamic excimer~\cite{Winnik}, respectively forming faster and slower than our time resolution (FWHM of the instrument response in 6~ns).

\section{Conclusion}

This study considered four pyrene-doped polystyrene coated acrylic samples which had various  concentrations and purities of pyrene. The analysis of these samples consisted of a characterization of their fluorescent response to UV light from 300~K to 4~K, in terms of the light yield and decay times.  At temperatures from 300~K to 4~K, depending on the pyrene sample, the light yield increased between 30\% to 45\%. This meant that the relative light yield at 4~K compared to evaporated TPB was at least 40\%. 
The decay profile of pyrene was modelled by two exponential terms: a short decay between 120~ns and 160~ns at 4~K, and a long decay between 300~ns and 350~ns at 4~K. All time constants increased when cooling down and are significantly longer than the nanosecond scale TPB decay.  At fixed purity, the highest pyrene concentration sample gave the highest light yield. At fixed concentration, the highest pyrene purity sample gave the highest light yield. The increase in purity was also found to lead to a higher contribution from the longer time constant decay to the overall decay at all temperatures.

One of the samples was examined in more detail by monitoring the monomer and excimer components separately using filters. The excimer dominates the light yield above $\sim 87$~K, the monomer dominates below that temperature.  The overall behaviour is qualitatively consistent with a reduction in non-radiative channels at low temperature. For the excimer, below 87~K, the results are consistent with a drop in exciton migration.

These pyrene-polystyrene films were developed  to act as wavelength shifters with slow time constants and high light yield. When these coatings are placed on specific detector components, these properties of pyrene films can help to provide additional pulseshape discrimination. Such films have been selected by the DEAP collaboration to deal with surface $\alpha$ events in certain regions of the liquid argon detector~\cite{gallacher_2021_arxiv}. Since the surface $\alpha$ background is a limiting background, the ability to discriminate these events increases the discovery potential of the detector. The high light yield and slow decays over the 4--300~K temperature range also makes these pyrene-polystyrene films applicable to background mitigation or event type identification in other liquid argon detectors, in other noble liquid and noble gas-based detectors, and, more generally, in liquid scintillator detectors.

\section{Acknowledgments}
Funding in Canada has been provided by NSERC through SAPPJ grants, by CFI-LOF and ORF-SIF, and by the McDonald Institute.
M.K. is supported by the International Research Agenda Programme AstroCeNT (MAB\allowbreak/2018\allowbreak/7) funded by the Foundation for Polish Science (FNP) from the European Regional Development Fund and  by the EU’s Horizon 2020 research and innovation program under grant agreement No~962480 (DarkWave).
Dr.~T.~R.~Pollmann from NIKHEF and TU M\"unchen provided comments on a draft of this work. Dr.~M.~Hamel of CEA Saclay provided insight into the properties of pyrene.  Queen's NSERC USRA summer student D.~Garrow contributed to the characterization of the instrument response.

\bibliographystyle{unsrt}

\bibliography{references}

\clearpage

\appendix

\section{\label{app:integral non exp}Calculation of monomer pulse integral}

The integral of the non-exponential pulse shape (Eq.~\ref{eq:LY_mono}) can be obtained as follows:
\begin{equation}
\begin{aligned}
\int_0^{+\infty} i_m(t)*E(t) dt 
& =\frac{N_1'}{\tau_1'} \int_{0}^{+\infty}   e^{-\tfrac{t}{\tau_1'}-2q\sqrt{\tfrac{t}{\tau_1'}}}~dt\\
& = 2 N_1'  \int_{0}^{+\infty} x e^{-x^2-2 x q} ~dx \mbox{ with } x=\sqrt{t/\tau_1'}\\
& = 2 N_1'  \left(\frac{1}{2} - q\int_{0}^{+\infty}  e^{-x^2-2 x q} ~dx \right) \mbox{ integrating by parts}\\
& =2 N_1' \left(\frac{1}{2} - q e^{q^2} \int_{0}^{+\infty} e^{-(x+q)^2}~dx \right)\\
& =N_1'\left(1- q e^{q^2}\sqrt{\pi} \left[ 1-{\mbox{erf}}(q) \right] \right) \mbox{ after } u=x+q\\
\end{aligned}
\end{equation}
where we have  used the  error function: $\mbox{erf}(x)=\frac{2}{\sqrt{\pi}} \displaystyle\int_{0}^{x} e^{-t^2}~dt $

\end{document}